\documentclass[preprint,fleqn,showpacs,showkeys]{revtex4}
\usepackage{graphicx}
\usepackage{amssymb}
\usepackage{amsmath}
\usepackage{bm}

\begin{document}
\setcounter{page}{1}

\title[]{Josephson Effect in Singlet Superconductor-Ferromagnet-Triplet Superconductor Junction}

\author{Chi-Hoon \surname{Choi}}

\affiliation{Department of Nanophysics, Gachon University, Seongnam 13120, Korea}

\begin{abstract}
We study the current-phase relation of a ballistic SIFIT junction, consisting of a spin-singlet superconductor (S), a weak ferromagnetic metal (F), a spin-triplet superconductor (T), and insulating ferromagnetic interfaces (I).
We use the generalized quasiclassical formalism developed by A. Millis \textit{et al.} to compute the current density and the free energy of the junction for arbitrary orientation of the magnetizations of the junction barrier. 
We investigate in detail the effect of the distribution of magnetization on the various harmonics of the current-phase relation and the transition of the ground state of the junction. 
The $\phi$-state junction can be realized for a noncollinear orientation of the barrier magnetizations in the plane perpendicular to the d-vector of the triplet superconductor. 
\end{abstract}

\pacs{74.50.+r, 74.20.Rp, 74.45.+c }
\keywords{Josephson effect, Superconductor-ferromagnet junction, Spin-triplet superconductor}

\maketitle

\section{INTRODUCTION}

The superconductor-ferromagnet junctions have been studied extensively both experimentally and theoretically.\cite{r1a,r1b,r1c} 
A ferromagnetic tunneling barrier can have a profound effect on the Josephson current-phase relation (CPR).\cite{r2}
The spin-singlet superconductor junctions with a nonmagnetic interface have the usual form of CPR, $I_S(\phi)=I_c \sin\phi$, where $I_c$ is the critical current and $\phi$ is the phase difference of the superconductors.
When the tunneling barrier has a ferromagnetic layer, the spin-triplet pairing amplitude can be induced in the ferromagnet and also the supercurrent can reverse its sign, making the $0-\pi$ transition.
Particularly, when the barrier has a multilayered ferromagnetic heterostructure with a nonuniform distribution of magnetization, several important features emerge.\cite{r3,r3a,r4}
The Josephson current can be dominated by the second harmonic $\sin 2\phi$ via the coherent transport of two Cooper pairs, which  makes it possible to realize the so-called $\phi$-junction.\cite{r3}
An anomalous supercurrent can also flow even for vanishing phase difference; $I_S(\phi =0) \neq 0$.\cite{r4}
These features of the dominant second harmonic, the $\phi$-junction, the anomalous Josephson effect (AJE) can play an important role in the development of the quantum electronic devices.\cite{r1c}

Recently, the Josephson junctions with spin-triplet superconductors have attracted much attention as the triplet superconductivity has been found in several materials such as $\rm{Sr_2 RuO_4}$ and the heavy fermion superconductors.\cite{r5a,r5b,r5c}
For a singlet superconductor-triplet superconductor junction, transport of a single Cooper pair is prohibited by symmetry and the supercurrent can tunnel through the barrier by the coherent transport of even numbers of Cooper pairs, leading to even harmonics $\sin( 2n\phi )$ in the CPR.
When the tunneling barrier of the junction has a magnetization, the $\cos\phi$-harmonic can appear because the transport of a single Cooper pair is possible due to a spin-flip scattering of the magnetic moment.\cite{r6a,r6a1,r6b,r6c}

In this paper, we study the current-phase relation of an SIFIT junction with a multilayered ferromagnetic heterostructure.
In the previous works on the singlet-ferromagnet-triplet junction, the tunneling barrier has been treated as a uniform ferromagnetic layer.\cite{r6a1,r6b,r6c}
The schematic diagram of the junction is depicted in the insert of Fig. 1. 
The interface ($\rm{I}$) is modeled by a delta-function like potential which can incorporate nonmagnetic as well as magnetic scatterings of quasiparticles.
The normal metal (N) of the middle layer can be replaced by a ferromagnetic metal. 
The singlet superconductor has an s-wave order parameter with the isotropic gap $\Delta_0$. 
For the triplet superconductor, we choose the following p-wave order parameter considered as a possible candidate for $\rm{Sr_2 RuO_4}$:   
\begin{equation}
\mathbf{d}(\mathbf{k})=\Delta_0 \hat {z} (k_x +ik_y ) e^{i\phi},
\end{equation}
where $\phi$ is the relative phase difference between the two superconductors.\cite{r5a} 

We compute the current density as a function of the phase difference in the ballistic limit while changing the magnetization of the interfaces and the exchange field of the ferromagnetic layer. 
We utilize the general formalism of A. Millis \textit{et al.} to take into account the interference effect due to scattering of quasiparticles from the neighboring interfaces.\cite{r7a} 
We focus on the question of how an inhomogeneous distribution of the magnetizations in the tunneling barrier affects the key features of the ferromagnetic Josephson junction such as the sign reversal of supercurrent, the $\phi$-junction, and the AJE.

Before going into details, we summarize our main results.
(i) For a wide range of the junction parameters, the current is well approximated by a combination of the harmonics of $\sin \phi$, $\cos\phi$, and $\sin 2\phi$.
Their relative magnitude is determined by the distribution of magnetizations in the tunneling barrier.
(ii) When the magnetizations are aligned with the d-vector of the triplet superconductor, the $\cos\phi$-term in the CPR appears, leading to the  AJE. 
(iii) When the magnetizations have a noncollinear distribution in the plane perpendicular to the d-vector, the $\sin \phi$-term can appear in addition to the second harmonic $\sin 2\phi$, leading to the $\phi$-junction. 
(iv) The ground state phase of the $\phi$-junction oscillates periodically as the strength of the exchange field changes due to the interference effect in the clean limit. 
(v) We also compute the pairing amplitude induced by the barrier magnetizations, which enables us to understand the key features of the CPR. 

\begin{figure}
\includegraphics[width=10 cm]{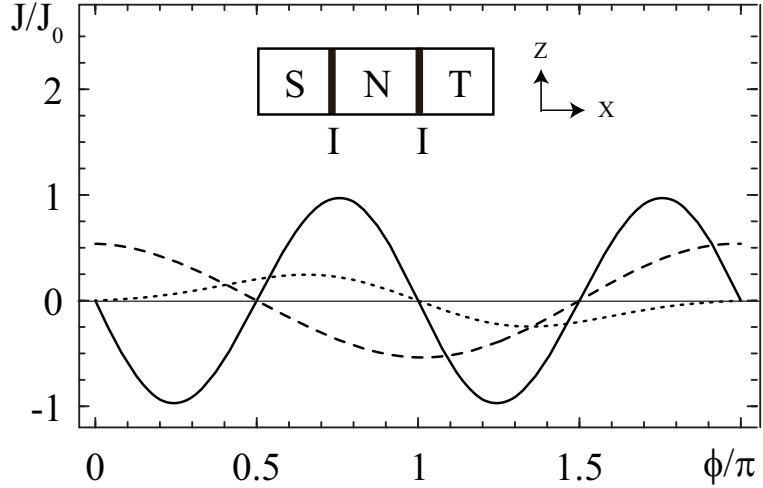}
\caption{Plots of the current density as a function of 
the phase difference for the SINIT junction.
The interface potential, defined by Eq. (5), is chosen as $\bar{U}_{I}=(0.5,\: 1)$, $(\hat{x},\: 4\hat{y})$, and $(2\hat{z},\: 2)$ for the solid, dotted, and dashed curves, respectively. 
The thickness of the normal layer is $\bar{d}=0.5$. 
In the insert, the SINIT junction with a normal metal layer surrounded by two half-infinite singlet and triplet superconductors is drawn.} 
\end{figure} 

\section{FORMALISM}

To do our calculation, we follow closely the formalism and the notations of Refs. 15 and 16.
We consider the SIFIT junction with the specular interfaces located at the positions of $x=0$ and $x=d$, as in Fig. 1.  
The Keldysh Green's function $\hat{G}(x,x')$ 
is decomposed into the four possible combinations of incoming and outgoing waves:  
\begin{equation}
\hat{\tau}_{3}\hat{G}^S(x,x')=\frac{1}{v^{S}}\underset{\alpha,\beta=\pm1}
{\sum}\hat{C}^{S}_{\alpha\beta}(x,x')e^{ip^{S}(\alpha x-\beta x')},
\end{equation}
where the superscript $S$ indicates each layer of the junction, and the subscripts $\alpha$ and $\beta$ are the indices representing the direction of momentum along the x-axis. 
The magnitude of the Fermi momentum along the x-axis is denoted by
$p^{S}$ and its corresponding Fermi velocity by $v^{S}=p^{S}/m$.

The interface is modeled by a delta-function like potential: 
$ \hat{v}=v_{0}\hat{1}+\mathbf{v}_{m}\cdot \boldsymbol{\sigma}$.
The charge and the magnetic scattering potentials of the interface are denoted by $v_{0}$ and $\mathbf{v}_{m}$, respectively.
The magnetic potential is assumed to be proportional to the magnetization $\mathbf{m}$ of the ferromagnetic interface.
The ferromagnetic metal is modeled by a weak exchange field and its potential in the particle space can be written as  
$\hat {V} =\mathbf{h} \cdot \boldsymbol{\sigma}$.\cite{r7c}

The differential equations and the boundary conditions for the amplitude function $\hat{C}_{\alpha \beta}$ are given by Eqs. (22-23) and Eq. (43) of Ref. 15, respectively.
The Green's function can be obtained by solving the differential equations for $\hat{C}_{\alpha\beta}$ with the proper boundary conditions at the interfaces at $x=0$ and $x=d$.\cite{r7b,r7d} 

Once the Green's function is computed, one can calculate the various physical quantities such as the current density, the pairing amplitude, and the density of states.
For our translationally-invariant planar interfaces, the current flows along the x-axis.
The current density from the particles incident with the momentum $\mathbf{p}$ can be calculated by
\begin{equation}
J(\hat{\mathbf{p}})=\frac{\pi}{2}N_{f}v_{f}
T\underset{n \geq 0}{\sum} (\hat{\mathbf{x}}\cdot\hat{\mathbf{p}}) \mathrm{Tr}
[\hat{\tau_{3}}(\hat{C}_{++} - \hat{C}_{- -} )], 
\end{equation}
where $N_{f}$ is the density of states at the Fermi energy.
The total current density can be computed by integrating the current density $J(\hat{\mathbf{p}})$ over the Fermi surface with $\hat{\mathbf{p}} \cdot \hat{\mathbf{x}} >0$.
The current is continuous at the interface due to particle conservation. 
The free energy of the junction can be obtained from the current density: 
\begin{equation}
E(\phi)=\frac{\Phi_{0}}{2\pi}\int_{0}^{\phi}J(\chi)d\chi,
\end{equation}
where $\Phi_{0}$ is the flux quantum.\cite{r2} 

\section{RESULTS AND DISCUSSION }

We now present our numerical calculations. 
We compute the current density of the SIFIT junction while varying the junction parameters such as the interface magnetizations, the exchange field, and the interlayer thickness. 
We assume that the Fermi velocity $v_f$ is the same everywhere and the singlet and the triplet superconductors have the same uniform gap $\Delta_{0}$ in each layer.
To simplify our notations, a variable $U_{I}$ is introduced for the interface potential:  
\begin{equation}
U_{I}=(U^L,\: U^R)=(v_{0}^L+\mathbf{v}_{m}^L,\: v_{0}^R+\mathbf{v}_{m}^R), 
\end{equation} 
where the superscripts R and L denote the right and the left interfaces. 
The energy and the length are scaled in units
of the superconducting gap $\Delta_{0}$ and the superconducting coherence length $\xi=\hbar v_{f}/\Delta_{0}$.
The following dimensionless quantities are defined: 
the interlayer thickness $\bar{d}=d/\xi$, the Fermi wave vector 
$\bar{k}_{f}=k_{f}\xi$, the interface potential 
$\bar{U}_{I}=U_{I}/(\hbar v_{f})$, 
and the exchange field $\bar{\mathbf{h}}=\mathbf{h}/\Delta_{0}$. 
In our calculation, we set $\bar{k}_{f}=1000$ and the temperature $T=0.1 \Delta_{0}$.

We compute the current density for a normal incidence by using Eq. (3).
The current density is normalized by $J_{0}=N_{f}v_{f} T/4$. 
For a different angle of incidence $\theta_k$, it is straightforward to generalize the calculations by replacing the position variable $x$ with $x/\cos\theta_k$.

We now discuss briefly the effect of the magnetization in the tunneling barrier on the pairing amplitude to understand its effect on the CPR.
It is summarized in Eq. (14) of Ref. 16 how the interface scattering potential can induce the various components of the pairing amplitude from its adjacent superconductor.
The magnetic potential $\mathbf{v}_{m}$ can induce the singlet pairing amplitude through the interaction term `$i \mathbf{v}_{m} \cdot \mathbf{f}$' and the triplet pairing amplitude through the terms of `$i \mathbf{v}_{m} f_0$' and `$i v_0 \mathbf{v}_{m} \times \mathbf{f}$', where $f_0$ and $ \mathbf{f}$ are the singlet and triplet pairing amplitudes of the superconductor, respectively. The induced pairing amplitudes acquire a $90^0$ phase shift.

In a similar way, one can derive the expression for the induced pairing amplitude by the exchange field.
We need to solve the differential equation for $\hat{C}_{++}$ in the ferromagnetic layer of the superconductor-ferromagnet junction while assuming that particles are incident from the superconductor. 
Up to the first order in $h=|\mathbf{h}|$, the transmitted pairing amplitude in the ferromagnetic layer can be written as
\begin{align}
f_0^{tr}&=e^{-2 \epsilon_n x /\hbar v_f} [\cos (qx) f_0 - i\sin (qx) \bar{\mathbf{h}} \cdot \mathbf{f}],
\nonumber \\
\mathbf{f}^{tr}&=e^{-2 \epsilon_n x/\hbar v_f} [ \mathbf{f} - i \sin (qx) \bar{\mathbf{h}} f_0],
\end{align}
where $q=2h/\hbar v_f$ and $x$ is the distance from the superconductor.
The exchange field $\mathbf{h}$ can thus induce the pairing amplitudes in a similar way to the interface magnetization, and it can be regarded as the interface potential with $v_0$=0 and $ \mathbf{v}_{m}  \propto e^{-2 \epsilon_n x/\hbar v_f} \sin (qx) \mathbf{h}$.
 
We investigated in detail the CPR of the SIFIT junction by calculating the current density as a function of the phase difference while changing the various junction parameters such as the magnitude and orientation of the interface potential at each interface and the exchange field as well as the interlayer thickness. 
Several observations are in order.
(i) It is found that  for a wide range of the junction parameters the current density can be approximated quite well by the following Fourier series: 
\begin{equation}
J(\phi)=C_1 \sin\phi + C_2 \cos\phi + C_3 \sin 2\phi. 
\end{equation}
(ii) The singlet-triplet junction has genetically the second harmonic $\sin2 \phi$ regardless of the tunneling barrier configurations.
(iii) There appears the $\cos\phi$-term in the CPR, accompanying the AJE, when the exchange field or the interface magnetization has a component parallel to the d-vector.
(iv) The $\sin\phi$-term appears, and thus the $\phi$-junction can be realized when the barrier magnetizations have a noncollinear distribution in the plane perpendicular to the d-vector. Or, it can be written as $\mathbf{m}_L \times \mathbf{m}_R \cdot \mathbf{d} \neq 0$, where $\mathbf{d}$ is the d-vector of the triplet superconductor and the $\mathbf{m}_L$ and $\mathbf{m}_R$ are the magnetizations of the left and right interfaces. 

We note that a single layer of magnetization is enough for the AJE, but at least two noncollinear magnetizations are required for the $\phi$-junction. This is in a sharp contrast to the singlet-ferromagnet-singlet junction in which case three non-coplanar magnetizations are needed for both AJE and $\phi$-junction such that $\mathbf{m}_L \times \mathbf{m}_R \cdot \mathbf{h} \neq 0$. \cite{r4}
Note also that the conditions for the AJE and $\phi$-junction in the singlet-ferromagnet-triplet junction are quite different from those in the triplet-ferromagnet-triplet junction. \cite{r7b} For the latter junction with the same triplet order parameters at both sides, the AJE and the $\phi$-junction occur at the same time as the d-vector of the triplet superconductor has both parallel and perpendicular components to a plane formed by the barrier magnetizations, requiring at least two noncollinear magnetizations. However, when the d-vectors at both sides of the junction are orthogonal to each other, the junction can have a generic $\sin 2\phi$-term in the CPR, and the AJE and the $\phi$-junction arise when one of the barrier magnetizations have a perpendicular component to the plane spanned by both d-vectors. 
The same $\cos\phi$-harmonic is responsible for both AJE and $\phi$-junction in the triplet-ferromagnet-triplet junction, while different harmonics are needed in the singlet-ferromagnet-triplet junction, i.e., the $\sin\phi$-harmonic for the $\phi$-junction and the $\cos\phi$-harmonic for the AJE.

For the singlet-ferromagnet-triplet junction, there always exists the second harmonic $\sin2\phi$ because the singlet and the triplet pairing amplitudes of the superconducting layers, $f_0$ and $f_z$ are orthogonal to each other. 
To have the $\cos\phi$-term, we need an interface magnetization or an exchange field parallel to the z-axis. 
The magnetization along the z-axis can induce the triplet pairing amplitude of $f_z$ from the singlet superconductor via the above-mentioned term `$i \mathbf{v}_{m} f_0$'. The same magnetization can also induce the singlet pairing amplitude from the triplet superconductor via `$i \mathbf{v}_{m} \cdot \mathbf{f}$'.
A Cooper pair tunneling between the induced singlet pairing amplitude and the singlet superconductor or the tunneling between the induced triplet pairing amplitude and the triplet superconductor, whose phases differ by $90^0$, leads to the $\cos\phi$-harmonic.
To have the $\sin\phi$-term, we need two separate magnetizations having components along the x- and y-axes.
For example, the pairing amplitude of $f_x$ can be induced both by the magnetization along the x-axis from the singlet superconductor via `$i \mathbf{v}_{m} f_0$' and by the magnetization along the y-axis from the triplet superconductor via `$i \mathbf{v}_{m} \times \mathbf{f}$'.
A Cooper pair tunneling between the induced pairing amplitudes from both sides leads to the $\sin\phi$-harmonic. 

First, we present the results for the SINIT junction whose tunneling barrier is composed of a normal metal layer and two interfaces. 
It is found that the dominant harmonics can be $\sin\phi$, $\cos\phi$, or $\sin2\phi$, depending on the orientation of the interface magnetizations. 
Three representative cases are plotted in Fig. 1. 
The Josephson current is dominated by the second harmonic $\sin 2\phi$ when both interfaces are nonmagnetic, as shown in the solid curve. 
 When the interface magnetization is aligned with the z-axis, as in the dashed curve, the leading harmonic can be the $\cos\phi$ with the anomalous Josephson current $J(\phi=0)\neq0$. 
The current is dominated by the $\sin\phi$-harmonic when the two interface magnetizations are aligned to the x- and y-axes, as in the dotted curve.

\begin{figure}
\includegraphics[width=10 cm]{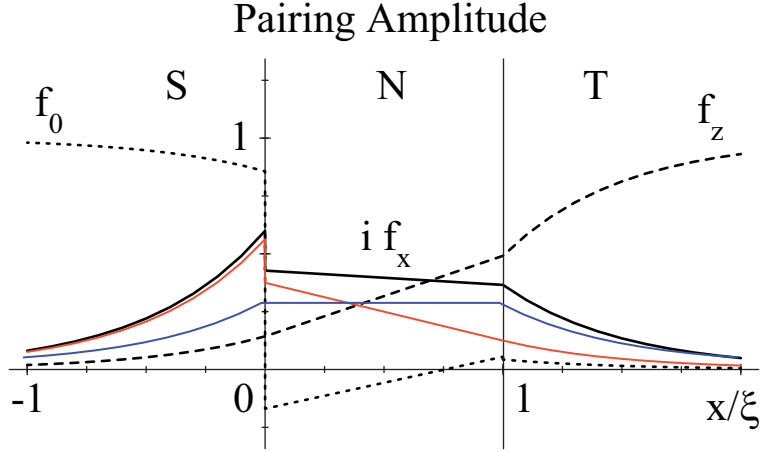}
\caption{(Color online) Plots of the pairing amplitudes as a function of the position for the SINIT junction.
The interface potential is chosen as $\bar{U}_{I}=(\hat{x},\: 1+\hat{y})$.
The red and the blue curves are the pairing amplitude $f_x$ induced by the left interface with $\bar{U}_{I}=(\hat{x},\: 0)$ and by the right interface with $\bar{U}_{I}=(0,\: 1+\hat{y})$, respectively.
The induced $f_x$'s are purely imaginary while $f_0$ and $f_z$ are real. 
The pairing amplitude is normalized by that of the bulk superconductor and we set $\bar{d}=1$, $\phi = 0 $, and $\bar{\epsilon}_n = 0.1 \pi$.} 
\end{figure} 

In Fig. 2, we plot the pairing amplitude induced by the interface magnetizations for the $\rm{SINIT}$ junction to understand its effect on the CPR.
We pay a particular attention to the case where the $\sin\phi$-harmonic is generated.
The interface parameter is chosen as $\bar{U}_{I}=(\hat{x},\: 1+\hat{y})$ so that the current is dominated by the $\sin\phi$:  $J/J_0 =2.03 \sin\phi -0.46 \sin 2 \phi$.
The pairing amplitude can be computed from $\hat{C}_{\alpha \beta}: $\cite{r7c}
\begin{equation}
f_n = \frac{\pi}{4} {\rm{Tr}} [( \hat{C}_{++} \pm \hat{C}_{- -} ) ( \hat{\tau}_1 -i  \hat{\tau}_2 ) (-i  \hat{\sigma}_2 ) \hat{\sigma}_n ], 
\end{equation}
where the upper sign corresponds to the singlet component, n=0, and the lower sign to the triplet component, n = x, y, and z.
The pairing amplitudes $f_0$ and $f_z$ have real values and decay in the normal metal layer. 
As discussed in the above, the pairing amplitude $f_x$ can be induced both by the magnetization along the x-axis at the left interface (the red curve) and by the magnetization along the y-axis at the right interface (the blue curve). The pairing amplitude $f_x$ induced by both sides (the solid black curve) is purely imaginary and has a fairly large value in the normal metal layer, leading to the dominant $\sin \phi$-harmonic. 

\begin{figure}
\includegraphics[width=10 cm]{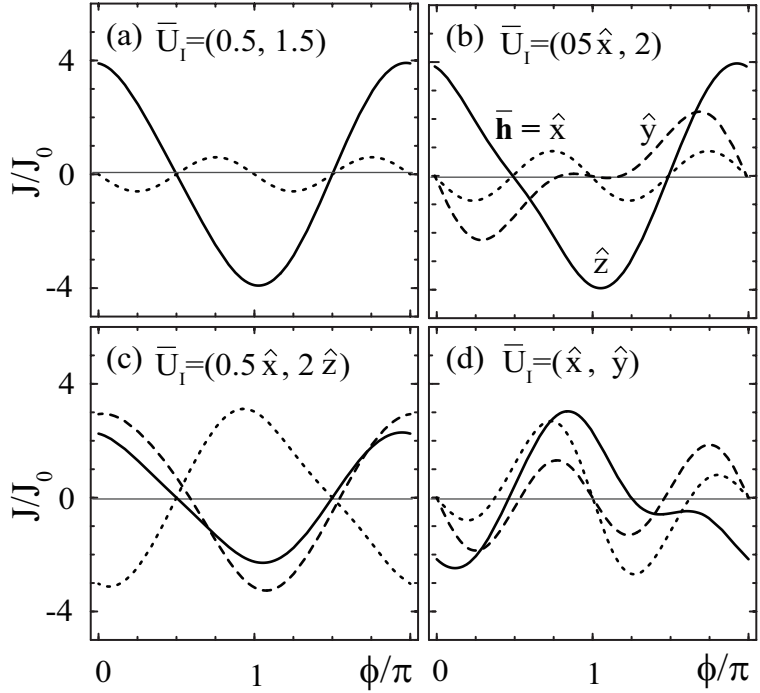}
\caption{Plots of the current density as a function of the phase difference for the SIFIT junction with several different sets of the interface potential $\bar{U}_{I}$.
The exchange field $\mathbf{h} $ is chosen to be parallel to the x-(dotted curve), y-(dashed), and z-(solid) axes. The interlayer thickness is $\bar{d}=0.5$.}
\end{figure} 

We now extend our discussion to the SIFIT junction to study the effect of the ferromagnetic layer on the Josephson current.
In Fig. 3, we present our numerical calculations showing the characteristic features of the singlet-ferromagnet-triplet junction.
The current is computed as a function of the phase difference for several typical types of the interface potentials with the exchange field along the x-, y-, and z-axes. 
The AJE occurs due to the $\cos\phi$-term whenever the interface magnetization or the exchange field has a z-component. 
In Fig. 3(a), where both interfaces are nonmagnetic, the current has only the second harmonic $\sin2\phi$ as long as the exchange field has no z-component.
In Fig. 3(b), where one of the interface magnetizations is along the x-axis, or in a direction perpendicular to the d-vector of the triplet superconductor, the current is still dominated by the $\sin 2\phi$ for the exchange field parallel to the same x-axis.
However, it undergoes a large change when the exchange field is rotated to the y-axis, due to the inclusion of the $\sin\phi$-term in the CPR. 
In Fig. 3(c), where the interface has a magnetization parallel to the d-vector of the triplet superconductor, the current is dominated by the $\cos\phi$-term independent of the orientation of the exchange field.
In Fig. 3(d), the $\phi$-junction is realized independent of the exchange field because the interface magnetizations along the x- and y-axes can generate the $\sin\phi$-term. The inclusion of the exchange field along the z-axis leads to the $\cos\phi$-term, which can make the magnitudes of all the three harmonics of $\sin\phi$, $\cos\phi$, and $\sin2\phi$ in the CPR become comparable.

\begin{figure}
\includegraphics[width=10 cm]{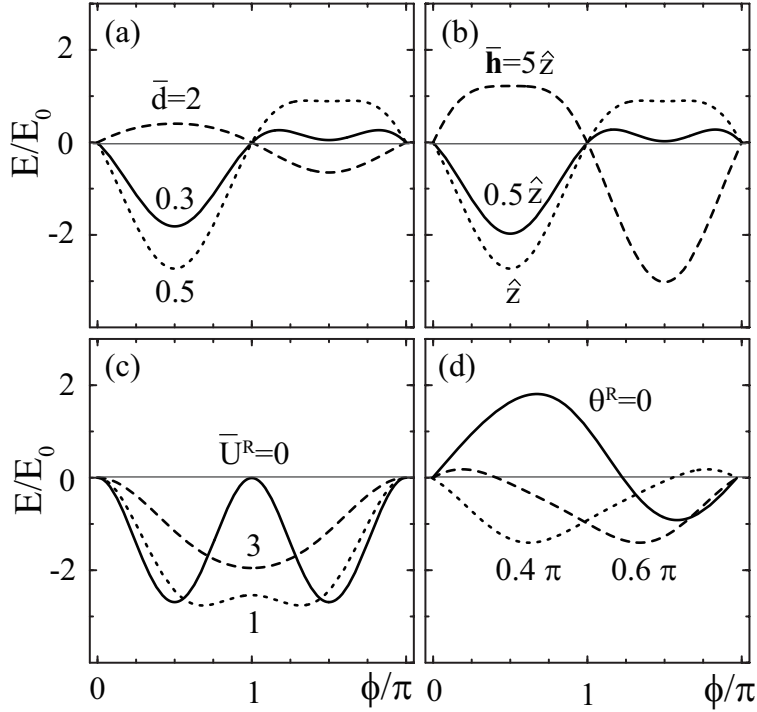}
\caption{Plots of the free energy as a function of the phase difference for the SIFIT junction with several different sets of the junction parameters.
In (a), different values of the interlayer thickness $\bar{d}$ are chosen for $\bar{\mathbf{h}}=\hat{z}$ and $\bar{U}_{I}=(\hat x,\: 1)$. 
In (b), different values of the exchange field $\bar{\mathbf{h}}$ are chosen for $\bar{d}=0.5$ and $\bar{U}_{I}=(\hat x,\: 1)$. 
In (c) and (d), different values of the nonmagnetic scattering potential $\bar{U}^{R}$ and the orientation angle of the magnetization $\theta^R$ at the right interface are chosen with the same set of the parameters; $\bar{d}=0.5$, $\bar{\mathbf{h}}=\hat{y}$, and $\bar{U}^{L}=\hat x$.}
\end{figure} 

Next, we discuss in detail the transition of the ground state by computing the free energy of the junction under the various situations.
The CPR of the form of Eq. (7) can, in general, show the AJE and the changes of ground state such as the $\phi$-junction and the $\pi/2-3\pi/3$ transition, depending on the signs and the relative magnitude of its coefficients. 
In Fig. 4, we plot the free energy as a function of the phase difference in the unit of $E_{0} =\Phi_{0}/{2\pi}$.
In Figs. 4(a) and 4(b), where either the exchange field or the interface magnetization has a z-component, the $\cos\phi$-term appears and the current is determined by the $C_2$ and $C_3$ terms: $J(\phi)= C_2 \cos\phi + C_3 \sin 2\phi$. 
Because the coefficient $C_3$ is negative in our parameter ranges,
the minimum of the free energy occurs at $\phi=3\pi/2$ for $C_{2}>0$ and at $\phi=\pi/2$ for $C_{2}<0$. 
Thus, the ground state makes a discrete transition from the $\pi/2$-state to the $3\pi/2$-state as the sign of $C_2$ changes.
This kind of the transition happens as the interlayer thickness changes as in Fig. 4(a), or as the strength of the exchange field changes as in Fig 4(b).

In Fig. 4(c), the exchange field is aligned with the y-axis and the magnetization of the left interface with the x-axis, so the $\sin\phi$-term appears and the current can be approximated by $J(\phi)=C_1 \sin\phi + C_3 \sin 2\phi$.
Because $C_{3}$ is negative, the free energy has double minima at the phases of 
\begin{equation}
\phi_{0}=\pm \cos^{-1}(-\frac{C_1}{2C_3}).
\end{equation}
The ground state can make a continuous transition to the $\phi_0$-state as the interface potential changes, as shown in Fig 4(c).
For example, $\phi_{0}=\pm 0.5 \pi$ for $\bar{U}^R=0$, $\phi_{0}=\pm 0.68 \pi$ for $\bar{U}^R=1$, and $\phi_{0}=\pi$ for $\bar{U}^R=3$. 

In Fig. 4(d), the free energy is computed for different orientations of the magnetization at the right interface.
The orientation angle $\theta^{R}$ is defined by an angle between the z-axis and the magnetization in the x-z plane. 
When the magnetization is along the z-axis such that $\bar{U}^R=\hat z$ ($\theta^R =0$), the $\cos\phi$-term in Eq. (7) is dominant. 
As the magnetization is inclined away from the z-axis, such as $\theta^R =0.4 \pi$, the $\cos\phi$-term weakens and the $\sin2\phi$-term becomes larger. This makes the magnitude of all the three coefficients in Eq. (7) become comparable. 
When the z-component of the magnetization is converted to a negative one, such as from $\theta^R =0.4 \pi$ to $0.6 \pi$, the sign of $C_2$ is reversed. 

\begin{figure}
\includegraphics[width=10 cm]{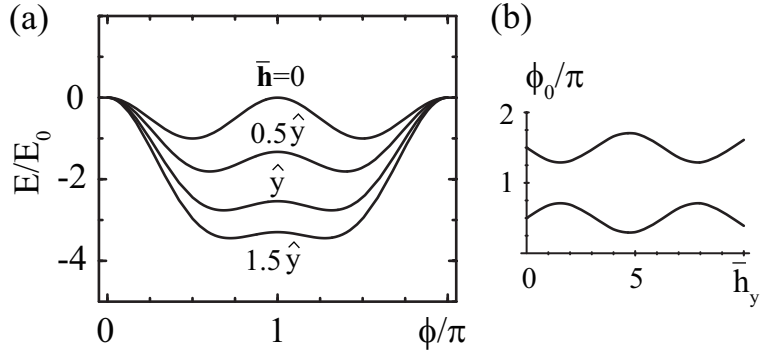}
\caption{(a) Similar plots of the free energy of the SIFIT junction as in Fig. 4 for different values of the exchange field along the y-axis, $\bar{h}_{y}$. We choose $\bar{d}=0.5$ and $\bar{U}_{I}=(\hat x,\: 1)$. (b) The ground state phase $\phi_{0}$ is plotted as a function of $\bar{h}_{y}$. It oscillates with the period of $2\pi$.}
\end{figure} 

In Fig. 5, we study the $\phi$-junction in more detail by calculating the free energy for different values of the exchange field.
The interface magnetization and the exchange field are aligned with the x- and y-axes, respectively, so that the current has the following form: $J(\phi)=C_1 \sin\phi + C_3 \sin 2\phi$. 
The ground state has two minima and its ground state phase $\phi_0$ changes continuously according to Eq. (9) as the strength of the exchange field changes. 
In Fig. 5(b), the ground state phase is plotted as a function of the strength of exchange field. 
The value of $\phi_0$ oscillates periodically due to a resonant scattering of quasiparticles in the ferromagnetic layer between the surrounding interfaces. 
In the ballistic limit, quasiparticles acquire a phase factor of $e^{2 i hd/\hbar v_f}$ in the ferromagnetic layer during the process of a Cooper pair tunneling.\cite{r1a}
This leads to the oscillation of the period of $2\pi$ in the Fig. 5(b).

In conclusion, we study the current-phase relation of the SIFIT junction for the various configurations of the magnetizations of the tunneling barrier.
The AJE, the $\phi$-junction, the $\pi/2-3\pi/3$ transition, and other types of CPR with the leading harmonics of $\sin\phi$, $\cos\phi$, and $\sin2\phi$ can be made readily in the SIFIT junction by adjusting the junction parameters.
The conditions for the AJE and $\phi$-junction in the singlet-ferromagnet-triplet junction are quite different from those in the singlet-ferromagnet-singlet and the triplet-ferromagnet-triplet junctions. This can, in principle, play a crucial role in identifying symmetry of the triplet superconductors.\cite{r8b}
In the future, we plan to extend our work to include the effect of strong exchange field of the ferromagnet, different Fermi velocities in each layer, and other types of the triplet superconducting order parameters.

\end{document}